\documentclass[aps,prb,twocolumn,groupedaddress,amsfonts,showpacs]{revtex4}
\usepackage{graphicx,color}
\usepackage{bm}

\begin{document}
\title{Spin-orbit induced anisotropy in the tunneling
magnetoresistance of magnetic tunnel junctions}
\author{A. Matos-Abiague and J. Fabian}
\affiliation{Institute for Theoretical Physics, University of
Regensburg, 93040 Regensburg, Germany}
\date{\today}

\begin{abstract}
The effects of the spin-orbit interaction on the tunneling
magnetoresistance of magnetic tunnel junctions are investigated. A
model in which the experimentally observed two-fold symmetry of
the anisotropic tunneling magnetoresistance (TAMR) originates from
the interference between Dresselhaus and Bychkov-Rashba spin-orbit
couplings is formulated. Bias induced changes of the
Bychkov-Rashba spin-orbit coupling strength can result in an
inversion of the TAMR. The theoretical calculations are in good
agreement with the TAMR experimentally observed in epitaxial
Fe/GaAs/Au tunnel junctions.
\end{abstract}


\maketitle

The tunneling magnetoresistance (TMR) effect refers to
ferromagnet/insulator/ferromagnet heterojunctions, in which the
magnetoresistance exhibits a strong dependence of the relative
magnetization directions in the different ferromagnetic layers and
their spin polarizations.\cite{julliere,maekawa,slonczewski}
Because of this peculiarly strong asymmetric behavior of the
magnetoresistance, TMR devices find multiple uses ranging from
magnetic sensors to magnetic random access memory
applications.\cite{maekawa,zutic}

It came as a surprise that the tunneling magnetoresistance may
strongly depend on the absolute orientation of the in-plane
magnetization directions with respect to a fixed crystallographic
axis, as experimentally observed in
Refs.~\onlinecite{gould,ruster,saito}. The phenomenon was termed
tunneling anisotropic magnetoresistance (TAMR).\cite{gould,brey}
Even more intriguing is the observation of the TAMR effect in
heterojunctions such as (Ga,Mn)As/AlOx/Au\cite{gould} and
Fe/GaAs/Au\cite{moser} sandwiches, where only one of the layers is
magnetic, and for which the TMR effect is absent. It has been
recognized that the origin of the TAMR effect is related to the
spin-orbit interaction
(SOI).\cite{gould,ruster,saito,brey,moser,shick} However, the
nature and details of the underlying mechanism producing the TAMR
remains a puzzle. In fact, it has become clear that the
responsible mechanisms for the TAMR can be different in different
systems. The (Ga,Mn)As/alumina/Au heterojunctions\cite{gould} and
in (Ga,Mn)As nanoconstrictions\cite{giddings} has been associated
with the anisotropic density of states in the ferromagnet
(Ga,Mn)As and was theoretically modelled by introducing strain
effects. First-principle calculations for the case of an Fe(001)
surface have recently demonstrated the appearance of the TAMR
effect due to the shifting of the resonant surface band via Rashba
SOI when the magnetization direction changes.\cite{chantis}
Furthermore, it has recently been observed that the symmetry axis
of the two-fold symmetry of the TAMR in Fe/GaAs/Au heterojunctions
can be flipped by changing the bias voltage.\cite{moser}

Here we formulate the model proposed to explain the experimental
results of Ref.~\onlinecite{moser}, in which the two-fold symmetry
of the TAMR observed in epitaxial ferromagnet/semiconductor/normal
metal junctions originates from the interface-induced $C_{2v}$
symmetry of the SOI arising from the interference of Dresselhaus
and Bychkov-Rashba spin-orbit couplings. Such interference effects
has already been investigated in lateral transport in 2D electrons
systems.\cite{schliemann} The symmetry, which is imprinted in the
tunneling probability becomes apparent in the contact with a
magnetic moment.

%Thus, the tunneling probability becomes itself anisotropic and the
%anisotropy of the density of states appears {\it a posteriori}, as
%a consequence of the anisotropic transmission and reflection at
%the interfaces of the heterojunction.

Consider a ferromagnet/semiconductor/normal-metal tunnel
heterojunction. The semiconductor is assumed to lack bulk
inversion symmetry (zinc-blende semiconductors are typical
examples). The bulk inversion asymmetry of the semiconductor
together with the structure inversion asymmetry of the
heterojunction give rise to the
Dresselhaus\cite{dresselhaus,roessler,winkler} and
Bychkov-Rashba\cite{winkler} SOIs, respectively. The interference
of these two spin-orbit couplings leads to a net, anisotropic SOI
with a $C_{2v}$ symmetry which is transferred to the tunneling
magnetoresistance when the electrons pass through the
semiconductor barrier. The model Hamiltonian describing the
tunneling across the heterojunction reads
\begin{equation}\label{hamilt}
    H=H_{0}+H_{Z}+H_{BR}+H_{D}.
\end{equation}
Here
\begin{equation}\label{h0}
H_{0}=-\frac{\hbar^2}{2}\nabla\left[\frac{1}{m(z)}\nabla\right]+V(z),
\end{equation}
with $V(z)$ the conduction band profile defining the potential
barrier along the growth direction ($z=[001]$) of the
heterostructure. The electron effective mass $m(z)$ is assumed to
be $m = m_{c}$ in the central (semiconductor) region and
$m=m_{l}=m_{r} \approx {m_{0}}$ (here $m_{0}$ is the bare electron
mass) in the left (ferromagnetic) and right (normal metal) layers.

The spin splitting due to the exchange field in the ferromagnetic
layer is given by
\begin{equation}\label{zeeman}
    H_{Z}=-\frac{\Delta(z)}{2}
    \mathbf{n}\cdot \mbox{\boldmath$\sigma$}.
\end{equation}
Here $\Delta(z)$ represents the exchange energy,
$\mbox{\boldmath$\sigma$}$ is a vector whose components are the
Pauli matrices, and $\mathbf{n}$ is a unit vector defining the
spin quantization axis determined by the in-plane magnetization
direction in the ferromagnet. The Zeeman splitting in the
semiconductor and normal metal can be neglected.

The Dresselhaus SOI can be written as\cite{roessler,
winkler,perel,ganichev,wang}
\begin{equation}\label{dresselhaus}
    H_{D}=\frac{1}{\hbar}(\sigma_{x}p_{x}-\sigma_{y}p_{y})
    \frac{\partial}{\partial z}\left(\gamma(z)\frac{\partial}{\partial
    z}\right),
\end{equation}
where $x$ and $y$ correspond to the $[100]$ and $[010]$
directions, respectively. The Dresselhaus parameter $\gamma(z)$
has a finite value $\gamma$ in the semiconductor region, where the
inversion bulk inversion asymmetry is present, and vanishes
elsewhere. Note that because of the step-like spatial dependence
of $\gamma(z)$, the Dresselhaus SOI [Eq.~(\ref{dresselhaus})]
implicitly includes the interface and bulk
contributions.\cite{roessler}

The Bychkov-Rashba SOI due to the interface inversion asymmetry is
incorporated in the model through the term\cite{note1}
\begin{equation}\label{rashba}
H_{BR}=\frac{1}{\hbar}\sum_{i=l,r}\alpha_{i}(\sigma_{x}p_{y}-\sigma_{y}p_{
x})\delta(z-z_{i}),
\end{equation}
where, $\alpha_l$ ($\alpha_r$) denotes the SOI strength at the
left (right) interface $z_{l}=0$ ($z_{r}=d$). The Bychkov-Rashba
SOI contribution inside the semiconductor can be neglected here.

Assuming that the in-plane wave vector $\mathbf{k}_{\parallel}$ is
conserved throughout the heterostructure, one can decouple the
motion along the growth direction ($z$) from the other spatial
degrees of freedom. The $z$ component of the scattering states in
the left (ferromagnetic) region [eigenstates of the Hamiltonian
(\ref{hamilt})] with eigenenergy $E$ can be written as
\begin{equation}\label{scattL}
\Psi_{\sigma}^{(l)}=\frac{e^{ik_{\sigma}z}\chi_{\sigma}}{\sqrt{k_{\sigma}}
}+
    r_{\sigma,\sigma}e^{-ik_{\sigma}z}\chi_{\sigma}+
    r_{\sigma,-\sigma}e^{-ik_{-\sigma}z}\chi_{-\sigma},
\end{equation}
where $z \leq 0$, $\chi_{\sigma}$ represents a spinor
corresponding to a spin parallel ($\sigma = \uparrow$) or
antiparallel ($\sigma = \downarrow$) to the magnetization
direction defined by the vector $\mathbf{n}$, and $k_{\sigma}$ is
the corresponding $z$ component of the wave vector in the left
region. In the central (semiconductor) region ($0<z<d$) we have
\cite{perel,wang}
\begin{equation}\label{scattC}
    \Psi_{\sigma}^{(c)}=
\sum_{i=\pm}(A_{\sigma,i}e^{q_{i}z}+B_{\sigma,i}e^{-q_{i}z})\zeta_{i},
\end{equation}
where $q_{\pm}=(1\mp 2m_{c}\gamma
k_{\parallel}/\hbar^2)^{-1/2}q_{0}$ (with $q_{0}$ being the $z$
component of the wave vector in the barrier in the absence of SOI)
and $\zeta_{\pm}$ are spinors corresponding to spins parallel
($+$) and antiparallel ($-$) to the direction
$\mathbf{k_{\parallel}}\times \mathbf{z}$, which is the
quantization direction in the barrier. In the right (normal metal)
region ($z \geq 0$) the scattering states read
\begin{equation}\label{scattR}
    \Psi_{\sigma}^{(r)}=
    t_{\sigma,\sigma}e^{i\kappa_{\sigma}(z-d)}\chi_{\sigma}+
    t_{\sigma,-\sigma}e^{i\kappa_{-\sigma}(z-d)}\chi_{-\sigma},
\end{equation}
where $\kappa_{\sigma}$ is the corresponding $z$ component of the
wave vector in the right region. The expansion coefficients in
Eqs.~(\ref{scattL}) - (\ref{scattR}) can be found by applying
standard matching conditions at each interface.\cite{andrada,wang}
Once the wave function is determined, the particle transmissivity
can be calculated from the relation
\begin{equation}\label{trans}
T_{\sigma}(E,k_{\parallel})=\textrm{Re}[\kappa_{\sigma}|t_{\sigma,\sigma}|
^2+
    \kappa_{-\sigma}|t_{\sigma,-\sigma}|^2].
\end{equation}
The current flowing along the heterojunction then is
\begin{equation}\label{current}
    I=\sum_{\sigma =\uparrow,\downarrow}I_{\sigma},
\end{equation}
where
\begin{equation}\label{current}
    I_{\sigma}=\frac{e}{(2\pi)^3\hbar}\int
    dE
d^2k_{\parallel}T_{\sigma}(E,\mathbf{k}_{\parallel})[f_{l}(E)-f_{r}(E)],
\end{equation}
and $f_l(E)$ and $f_{r}(E)$ are the electron Fermi-Dirac
distributions with chemical potentials $\mu_{l}$ and $\mu_{r}$ in
the left and right leads, respectively.

The TAMR refers to the changes of the tunneling magnetoresistance
($R$) when varying the magnetization direction $\mathbf{n}$ of the
magnetic layer with respect to a fixed axis. Here we assume the
magnetoresistance $R_{[110]}$, measured when $\mathbf{n}$ points
in the [110] crystallographic direction as a reference. The TAMR
is then given by
\begin{equation}\label{tamr}
    \textrm{TAMR}_{[110]}(\theta)=\frac{R(\theta)-R_{[110]}}{R_{[110]}}=
    \frac{I_{[110]}-I(\theta)}{I(\theta)},
\end{equation}
where $\theta$ is the angle between the magnetization direction
$\mathbf{n}=(\cos\theta,\sin\theta,0)$ and the [100]
crystallographic axis. We also find it useful to define the
tunneling anisotropic spin polarization (TASP) as
\begin{equation}\label{tasp}
    \textrm{TASP}_{[110]}(\theta)=\frac{P_{[110]}-P(\theta)}{P(\theta)}.
\end{equation}
The TASP measures the changes in the tunneling spin
polarization\cite{maekawa,zutic,perel}
$P=(I_{\uparrow}-I_{\downarrow})/I$ (which is a measurable
quantity\cite{zutic} accounting for the polarization efficiency of
the transmission) when rotating the in-plane magnetization in the
ferromagnet.

For a concrete demonstration of the proposed theoretical model we
performed zero-temperature calculations of the TAMR in an
epitaxial Fe/GaAs/Au heterojunction similar to that used in the
experimental observations reported in Ref.~\onlinecite{moser}. We
use the value $m_{c}=0.067\;m_{0}$ for the electron effective mass
in the central (GaAs) region. The barrier width and high (measured
from the Fermi energy) are, respectively, $d=80\;\textrm{\AA}$ and
$V_{c}=0.75 \textrm{ eV}$, corresponding to the experimental
samples in Ref.~\onlinecite{moser}. For the Fe layer a Stoner
model with the majority and minority spin channels having Fermi
momenta $k_{F\uparrow}=1.05 \times 10^8 \textrm{ cm}^{-1}$ and
$k_{F\downarrow}=0.44 \times 10^8 \textrm{ cm}^{-1}$, \cite{jwang}
respectively, is assumed. The Fermi momentum in Au is taken as
$\kappa_{F}=1.2 \times 10^8 \textrm{ cm}^{-1}$. \cite{ashcroft} We
consider the case of relatively weak magnetic fields
(specifically, $B=0.5\textrm{ T}$). At high magnetic fields, say,
several Tesla, our model is invalid as it does not include
cyclotron effects relevant when the cyclotron radius becomes
comparable to the barrier width.

\begin{figure}
\includegraphics[width=6cm]{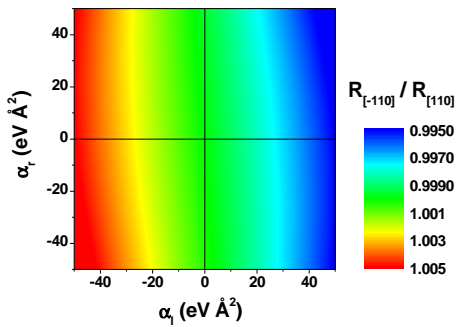}
\caption{Values of the ratio $R_{[\bar{1}10]}/R_{[110]}$ as a
function of the interface Bychkov-Rashba parameters $\alpha_{l}$
and $\alpha_{r}$.} \label{F1}
\end{figure}

\begin{figure}
\includegraphics[width=7cm]{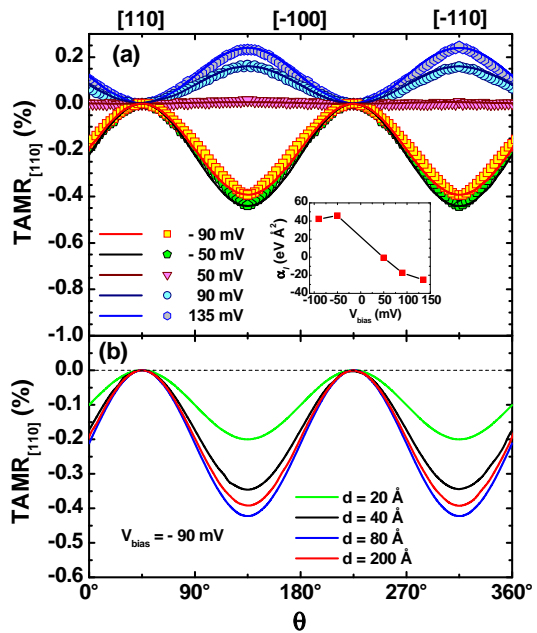}
\caption{(a) Angular dependence of the TAMR in an Fe/GaAs/Au
tunnel heterojunction for different values of the bias voltage
$V_{bias}$. Solid lines corresponds to our theoretical results
while symbols represent the experimental data (conveniently
mirrored) as deduced from Ref.~\onlinecite{moser}. The values of
the phenomenological parameter $\alpha_{l}$ have been determined
by fitting the theory to the experimental values of the ratio
$R_{[\bar{1}10]}/R_{[110]}$ for each value of $V_{bias}$. The
extracted bias dependence of $\alpha_{l}$ is shown in the inset.
(b) Angular dependence of the TAMR for different barrier
widths.}\label{F2}
\end{figure}

The Dresselhaus spin-orbit parameter in GaAs is $\gamma = 24
\textrm{ eV \AA}^{3}$.\cite{winkler,perel,ganichev} The values of
the Bychkov-Rashba parameters $\alpha_{l}$, $\alpha_{r}$ [see
Eq.~(\ref{rashba})] are not know for metal-semiconductor
interfaces. Due to the complexity of the problem, a theoretical
estimation of such parameters requires first principle
calculations including the band structure details of the involved
materials, which is beyond the scope of the present paper. Here we
assume $\alpha_{l}$ and $\alpha_{r}$ as phenomenological
parameters. In order to investigate how does the degree of
anisotropy depend on these two parameters we performed
calculations of the ratio $R_{[\bar{1}10]}/R_{[110]}$ (which is a
measure of the degree of anisotropy\cite{moser}) as a function of
$\alpha_{l}$ and $\alpha_{r}$. The results are shown in
Fig.~\ref{F1}, where one can appreciate that the size of this
ratio (and, consequently, of the TAMR) is dominated by
$\alpha_{l}$. Then, since the values of the TAMR are not very
sensitive to the changes of $\alpha_{r}$ we can set this
parameter, without loss of generality, to zero. This leaves
$\alpha_{l}$ as a single fitting parameter when comparing to
experiment. Such a comparison is shown in Fig.~\ref{F2}(a) for
different values of the bias voltage. The agreement between theory
and experiment is very satisfactory. The values of the
phenomenological parameter $\alpha_{l}$ are determined by fitting
the theory to the experimental value of the ratio
$R_{[\bar{1}10]}/R_{[110]}$ and this is enough for our theoretical
model to reproduce the {\it complete} angular dependence of the
TAMR, i.e., the proposed model is quite robust. Assuming that the
interface Bychkov-Rashba parameters are voltage
dependent\cite{zutic} (unlike $\gamma$, which is a material
parameter) we perform the same fitting procedure for the available
experimental data corresponding to different bias voltages and
extract the bias dependence of $\alpha_{l}$ [see the inset in
Fig.~\ref{F2}(a)], from which we estimate the value $\alpha_{l}
\approx 23 \textrm{ meV \AA}^{2}$ at zero bias. This value is
comparable to the corresponding value of the interface
Bychkov-Rashba parameter $\alpha_{l} \approx 27 \textrm{ meV
\AA}^{2}$ obtained from a five level $\mathbf{k}\cdot \mathbf{p}$
model \cite{zawadzki} for an InAs/GaAs interface. Selecting
InAs/GaAs for comparison with Fe/GaAs we only wish to show that
our fitted values have reasonable magnitude, not differing too
much from known values in semiconductor interfaces. Interestingly
$\alpha_{l}$ in our system changes sign at a bias slightly below
50 mV. This bias induced change of the interface Bychkov-Rashba
parameter results in an inversion of the TAMR [see
Fig.~\ref{F2}(a)]. Similar behavior is reported by ab initio
calculations on Fe surfaces, where only Bychkov-Rashba SOI is
present.\cite{chantis}

The robustness of our model can be understood from the following
simplified picture of the TAMR effect. The SOI term
$H_{SO}=H_{D}+H_{BR}$ can be written [see Eqs.~(\ref{dresselhaus})
and (\ref{rashba})] as a Zeeman-like term $H_{SO}\sim
\hat{\mathbf{B}}_{eff}\cdot\mbox{\boldmath$\sigma$}$ with the
effective magnetic field
\begin{equation}\label{beff}
    \hat{\mathbf{B}}_{eff}(\mathbf{k}_{\parallel})=(\alpha_{l}\delta(z)k_{y}-\gamma
k_{x}\partial_{z}^{2},-\alpha_{l}\delta(z)k_{x}+\gamma
k_{y}\partial_{z}^{2},0),
\end{equation}
where, for the sake of qualitative argument we neglect the
interface Dresselhaus contributions. Performing the average of
$\hat{\mathbf{B}}_{eff}$ over the unperturbed (in the absence of
SOI) states of the system one obtains the following general form
of the averaged effective spin-orbit magnetic field
\begin{equation}\label{www}
    \mathbf{w}(\mathbf{k}_{\parallel})=(\tilde{\alpha}_{l}
k_{y}-\tilde{\gamma} k_{x},-\tilde{\alpha}_{l}
k_{x}+\tilde{\gamma} k_{y},0),
\end{equation}
where $\tilde{\alpha}_{l}=\alpha_{l}f_{\alpha}(k_{\parallel})$ and
$\tilde{\gamma}=\gamma f_{\gamma}(k_{\parallel})$, with
$f_{\alpha}(k_{\parallel})$ and $f_{\gamma}(k_{\parallel})$ being
real functions of $k_{\parallel}=|\mathbf{k}_{\parallel}|$. The
effective field $\mathbf{w}(\mathbf{k}_{\parallel})$ becomes
anisotropic in the $\mathbf{k}_{\parallel}$-space with a $C_{2v}$
symmetry when both $\alpha_{l}$ and $\gamma$ have finite values.
It characterizes the amount of $\mathbf{k}_{\parallel}$-dependent
precession of the electron spin during the tunneling. For a given
$\mathbf{k}_{\parallel}$ there are only two preferential
directions in the system, defined by $\mathbf{n}$ and
$\mathbf{w}(\mathbf{k}_{\parallel})$. Therefore, the anisotropy of
a scalar quantity such as the total transmissivity
$T(E,\mathbf{k}_{\parallel})=T_{\uparrow}(E,\mathbf{k}_{\parallel})+
T_{\downarrow}(E,\mathbf{k}_{\parallel})$ can be obtained as a
perturbative expansion in powers of
$\mathbf{n}\cdot\mathbf{w}(\mathbf{k}_{\parallel})$, since the SOI
is much smaller than the other relevant energy scales in the
system. The total transmissivity is then given, up to second order
in the anisotropy, by $T(E,\mathbf{k}_{\parallel})\approx
T^{(0)}(E,k_{\parallel})+T^{(1)}(E,k_{\parallel})\mathbf{n}\cdot\mathbf{w}(\mathbf{k}_{\parallel})+
T^{(2)}(E,k_{\parallel})[\mathbf{n}\cdot\mathbf{w}(\mathbf{k}_{\parallel})]^{2}$.
Averaging over the in-plane momenta to get the full current, one
obtains for small bias voltages,
\begin{equation}\label{condani}
    I=\langle
T^{(0)}(E_{F},k_{\parallel})\rangle+\langle
T^{(2)}(E_{F},k_{\parallel})[\mathbf{n}\cdot\mathbf{w}(\mathbf{k}_{\parallel})]^{2}\rangle
V_{bias},
\end{equation}
where $\langle ...\rangle$ represents average over
$\mathbf{k}_{\parallel}$. Note that the first order term vanishes
after average over $\mathbf{k}_{\parallel}$ since
$\mathbf{w}(\mathbf{k}_{\parallel})=-\mathbf{w}(-\mathbf{k}_{\parallel})$.
Taking into account Eqs.~(\ref{tamr}), (\ref{www}), and
(\ref{condani}) one obtains the following approximate expression
for the TAMR
\begin{equation}\label{tamrani}
    \textrm{TAMR}_{[110]}\approx \frac{\langle
\tilde{\alpha}_{l}\tilde{\gamma}T^{(2)}k_{\parallel}^{2}\rangle
[\sin(2\theta)-1]}{\langle
T^{(0)}\rangle_{\mathbf{k}_{\parallel}}}\propto \alpha_{l}\gamma
    [\sin(2\theta)-1],
\end{equation}
where the arguments of the expansion coefficients $T^{(0)}$ and
$T^{(2)}$ have been omitted for brevity. The angular dependence in
Eq.~(\ref{tamrani}) is consistent with that found experimentally,
as well as that obtained from the full theoretical calculations
[see Fig.~\ref{F2}(a)]. One can clearly see from
Eq.~(\ref{tamrani}) that bias-induced changes of the sign of the
Bychkov-Rashba parameter $\alpha_{l}$ lead to an inversion of the
TAMR. When $\alpha_{l}\gamma = 0$, the two-fold TAMR is
suppressed. To put in words, the Bychkov-Rashba (or Dresselhaus)
term by itself cannot explain the observed $C_{2v}$ symmetry. The
suppression of the TAMR is approximately realized in
Fig.~\ref{F2}(a) for the case of a bias voltage of 50 mV.

The above model neglects the contribution of the
spin-orbit-induced symmetries of the involved bulk structures.
Say, Fe exhibits a four-fold anisotropy, which should be reflected
in the tunneling density of states. The fact that this is not seen
in the experiment suggests that this effect is smaller than the
two-fold symmetry considered in our model.
%\cite{note2}

Another system parameter that can influence the size of the TAMR
is the width of the barrier. The angular dependence of the TAMR
for the case of $V_{bias}=-90\textrm{ meV}$ is displayed in
Fig.~\ref{F2}(b) for different values of the barrier width $d$. As
clearly seen in Fig.~\ref{F2}(b), our model predicts an increase
of the TAMR when increasing the width of the barrier.

\begin{figure}
\includegraphics[width=6cm]{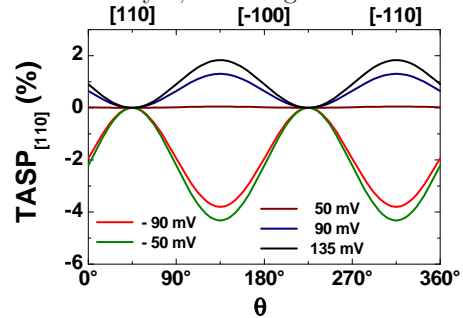}
\caption{Angular dependence of the TASP for different values of
the bias voltage.}\label{F3}
\end{figure}

Finally, we show the angular dependence of the TASP [see
Eq.~(\ref{tasp})] in Fig.~\ref{F3} for different values of the
bias voltage. The anisotropy of the tunneling spin polarization
indicates that the amount of transmitted and reflected spin at the
interfaces depends on the magnetization direction in the Fe layer,
resulting in an anisotropic spin local density of states at the
Fermi surface\cite{matos} and showing spin-valve-lake
characteristics.

In summary, we have formulated a theoretical model in which the
two-fold symmetry of both the TAMR and the TASP in epitaxial
ferromagnet/semiconductor/normal-metal heterojunctions originates
from the interplay between the Dresselhaus and Bychkov-Rashba
SOIs.
%The size and sign of the TAMR and TASP can be manipulated by
%bias induced changes in the interface Bychkov-Rashba spin-orbit
%coupling. Our model predicts an increase of the TAMR with
%increasing the barrier width. Explicit calculations were performed
%for the case of an epitaxial Fe/GaAs/Au heterojunction. The
Our theoretical results for epitaxial Fe/GaAs/Au heterojunctions
are in good agreement with the available experimental data.

This work was supported by the Deutsche Forschungsgemeinschaft via
SFB 689.

%\begin{references}

% \end{multicols}


\begin{thebibliography}{00}
\bibitem{note1} Eq.~(\ref{rashba}) can be well justified at
semiconductor interfaces.\cite{andrada} We propose it here as a
phenomenological model for metal/semiconductor interfaces, as the
simplest description of the interface-induced SOI symmetry. For
metalic surfaces the Bychkov-Rashba SOI has already been
investigated.\cite{henk} We assume that electrons with small
transverse momenta $p_{x}$ and $p_{y}$ have sizable tunneling
probabilities, justifying the linear character of the SOI.
\end{thebibliography}
\end{document}